\documentclass[12pt]{article}
\usepackage{graphicx}
\usepackage{ifpdf}
\ifpdf  \fi
\usepackage{amsfonts,amssymb}
\usepackage{amsthm}%% The amsthm package provides extended theorem environments
\usepackage{amsmath}
\usepackage[dvipsnames]{xcolor}
\usepackage[ruled,vlined]{algorithm2e} %% ADDS ALGORITHMS
\usepackage{newtxtext}

\everymath{\displaystyle}

\usepackage{subcaption}
\usepackage[colorinlistoftodos]{todonotes}

\usepackage{mathptmx,helvet,courier,makeidx,multicol,footmisc}
\usepackage[numbers]{natbib}
\bibpunct{(}{)}{;}{a}{,}{,}

\usepackage[bookmarks=false, pdfauthor={Irene Martinez}]{hyperref}
    \hypersetup{colorlinks,
      linkcolor=blue,
      citecolor=Emerald,
      urlcolor=blue}

\usepackage[colorinlistoftodos]{todonotes}

\usepackage{nomencl}
\makenomenclature

\newcommand{\commentout}[1]{}

\newcommand{\ba}{\begin{array}}
        \newcommand{\ea}{\end{array}}
\newcommand{\bc}{\begin{center}}
        \newcommand{\ec}{\end{center}}
\newcommand{\bdm}{\begin{displaymath}}
        \newcommand{\edm}{\end{displaymath}}
\newcommand{\bds} {\begin{description}}
        \newcommand{\eds} {\end{description}}%17Apr01
\newcommand{\ben}{\begin{enumerate}}
        \newcommand{\een}{\end{enumerate}}
\newcommand{\beq}{\begin{equation}}
        \newcommand{\eeq}{\end{equation}}
\newcommand{\bfg} {\begin{figure}[h]}
        \newcommand{\efg} {\end{figure}}%Nov 5,99
\newcommand{\bi} {\begin {itemize}}
        \newcommand{\ei} {\end {itemize}}
\newcommand{\bqn}{\begin{eqnarray}}
        \newcommand{\eqn}{\end{eqnarray}}
\newcommand{\bqs}{\begin{eqnarray*}}
        \newcommand{\eqs}{\end{eqnarray*}}
\newcommand{\bsl} {\begin{slide}[8.8in,6.7in]}
        \newcommand{\esl} {\end{slide}}
\newcommand{\bsq}{\begin{subequations}}
        \newcommand{\esq}{\end{subequations}}       
\newcommand{\bss} {\begin{slide*}[9.3in,6.7in]}
        \newcommand{\ess} {\end{slide*}}
\newcommand{\btb} {\begin {table}}
        \newcommand{\etb} {\end {table}}%Nov 10,99

\def\pmb#1{\setbox0=\hbox{$#1$}%
   \kern-.025em\copy0\kern-\wd0
   \kern.05em\copy0\kern-\wd0
   \kern-.025em\raise.0433em\box0 }

\usepackage[draft]{changes}
\definechangesauthor[name={IM},color=red]{IM}

\usepackage[colorinlistoftodos]{todonotes}

\graphicspath{ {figures/} }

\usepackage{multirow}
\usepackage{float}

\usepackage{setspace}   %Allows double spacing with the \doublespacing command

\begin{document}

\title{Mitigation of stop-and-go traffic waves with intelligent vehicles at low market penetration rates} %2018/11/15

\author{Irene Mart\'inez\footnote{Department of Transport \& Planning,  Technische Universiteit Delft. i.martinezjosemaria@tudelft.nl. Corresponding author.} }

\maketitle

\begin{abstract}

Stop-and-go traffic patterns sometimes manifest on roadways without any discernible congestion triggers. Such a phenomenon has been observed on homogeneous ring roads without lane changes. With the development of vehicle technology and measurement sensors, multiple researchers have focused on studying the influence of automated vehicles on traffic. In particular, there is a focus on the design of string-stable adaptive cruise control (ACC) strategies to dampen stop-and-go waves. However, there is no systematic comparison among different strategies nor a quantitative analysis of the oscillation reduction at low market penetration rates (MPRs). This paper proposes a framework to evaluate the impact of low MPRs across multiple ACC strategies. Then, through Monte Carlo simulations,  our findings indicate that multi-vehicle anticipation technology yields nearly equivalent benefits in mitigating stop-and-go patterns compared to full vehicular connectivity, even at a modest MPR of 1\%. In contrast, partial connectivity among vehicles only eliminates stop-and-go waves if the MPR is larger.

\end{abstract}


\newpage

%\doublespacing

\section{Introduction}\label{sec:intro}

Traffic congestion increases  travel times, the risk of traffic accidents, and energy consumption \citep{Li2014}.  
There are multiple well-studied congestion triggers, such as lane changes or road bottlenecks (geometrical changes that reduce the road capacity, e.g., a reduction in lanes). However, human drivers may spontaneously generate traffic congestion in the form of stop-and-go traffic in a homogeneous road stretch without any apparent reason \citep{Sugiyama2008Traffic, Tadaki2013, Stern2018}. This phenomenon is referred to as ``phantom jam'', and arises  when fairly small disturbance of speed ends up causing a complete stop of vehicles due to the amplification of perturbations.
Moreover, stop-and-go traffic has also been observed upstream of sag bottlenecks \citep{Koshi1992}.
Several reasons for the formation of stop-and-go have been explored in the transportation literature. Some researchers have tried to explain these traffic flow instabilities through macroscopic flow \citep{Flynn2009}, and microscopic flow  \citep{Bando1995} models. 
It has been proposed that a small perturbation may grow, leading to stop-and-go traffic because of string instabilities in mathematical models. Alternatively,  some researchers introduced several additional parameters to capture how heterogeneous driving behaviors can result in the generation and propagation of these traffic oscillations  \citep{Chen2012A}.

The stop-and-go phenomenon is closely related to research on the string stability of car-following models. However, most of the studies analyze the stability of car-following models through the propagation of disturbances across vehicles after externally perturbing (e.g., giving a `kick' to) the first or second vehicle of a platoon \citep{wilson2021car-following, Montanino2021On}. Alternatively, others have studied the speed profile of the following vehicles when the leader follows an oscillatory speed pattern \citep{Shang2021Impacts}. %However, by definition, 
\textcolor{black}{These models are not able to endogenously reproduce }the phantom jam phenomenon, which is formed without external disruption according to experimental results \citep{Sugiyama2008Traffic, Tadaki2013, Stern2018}. Few models are able to reproduce the stop-and-go phenomena without this external perturbation.
To do so, some researchers introduce sources of randomness  \citep{Chen2012A, Alcoba2019LWR}, e.g.,
by extending the  Newell's simplified car-following (NSCF) model \citep{Newell2002} to incorporate a random noise term. 
In the transportation field, the randomness related to the inherent system dynamics is mainly attributed to the driver’s behavior, which can be best described in distributional rather than deterministic parameters \citep{Li2008Investigation}.

Current technology and traffic management strategies are unable to dampen these stop-and-go waves. 
Fortunately, recent simulation-based studies \citep{Talebpour2016Influence, Ghiasi2019, Yu2021Automated} indicate that intelligent (i.e., automated and/or connected) vehicles have the ability to mitigate (or eliminate) stop-and-go if their longitudinal control is designed carefully.

Many researchers have studied the impact of automated driving on traffic flow \citep{Talebpour2016Influence, Yu2021Automated}.
Depending on the level of automation and connectivity, the vehicles can continuously monitor the distance to the car in front and other surrounding traffic. In particular, many commercially available vehicles already have adaptive cruise controller (ACC) devices that  adjust the acceleration following certain  rules, i.e., 
\begin{align}
    \dot v(t) = f (g(t), \dot g(t), v(t)), \label{eq/general}
\end{align}
where $g(t)$ is the gap (distance between the front bumper of the vehicle and the rear bumper of its leader), and $v(t)$ is the speed. In the traffic flow literature, speed (or acceleration) is usually defined as a function of the spacing $s(t)=g(t)+l$, where $l$ is the average length of the vehicles. In the future, vehicles are expected to have certain communication capabilities, through vehicle-to-vehicle (V2V) or vehicle-to-infrastructure (V2I) communications, and a cooperative adaptive cruise controller (CACC) could be used to leverage information from other vehicles or the infrastructure.

%Many researchers have investigated 
The impacts of a platoon of vehicles equipped with ACC and CACC capabilities on traffic flow stability have been extensively studied both analytically and through simulations. To do so, researchers test the impact of introducing multiple ACC-equipped vehicles in the traffic stream with variations on (\ref{eq/general}). Generally, human-driven vehicles (HVs) are assumed to behave following a traditional car-following model, e.g., the intelligent drivers model \citep{Treiber2000}, or optimal velocity model \citep{Bando1995}.
One of the most extended ACC strategies is the so-called constant time headway policy proposed in 1993 \citep{Ioannou1993Intelligent}. Similar to other researchers \citep{Shang2021Impacts},  we will refer to this policy as a constant time gap ($\tau$) to be consistent with the common traffic flow terminology\footnote{Time headway is defined as $h= \frac{s}{v}$ and time gap $\tau= \frac{s-l}{v}$, where $s$ is the spacing, $l$ the length of the vehicle, and $v$ the speed.}.
The constant time gap strategy aims to maintain a desired spacing based on the leaders' speed, i.e., $s_d = s_j + \tau v$, where $s_j$ is the jam spacing. It represents the minimum spacing when the leader is stopped ($v=0$). This strategy is equivalent to the NSCF model \citep{Newell2002} if the throttle and the brake can guarantee instantaneous speed adjustment.

However, there has been less attention in the literature towards mixed traffic scenarios with low market penetration rate (MPR) of vehicles equipped with ACC or CACC. It is essential to focus on these  more realistic and near-future traffic scenarios.
A first study in this direction analyzed the stability and controllability of a ring road with 20 vehicles and a single automated vehicle to control traffic oscillations \citep{Cui2017}. They showed that string stability could be achieved, but the effectiveness of such control is limited when the model has noise.
In the same line of stabilizing flow with a low MPR, a deep reinforcement learning trained with real vehicle trajectories was proposed to control the longitudinal movement of a single connected and automated vehicle (CAV) behind a HV, which is then followed by a platoon of HVs \citep{Jian2021}. The oscillations of the CAV were damped by 54\% while HVs oscillations were reduced by 8-28\%. 
Moreover, recent studies also explore the fact that a single AV could measure the state of multiple vehicles ahead \citep{Chen2009, Dona2022Multianticipation}. This ACC with multi-vehicle anticipation was shown to dissipate stop-and-go.

Further, there is a significant increase in experimental studies to evaluate the stability of the proposed ACC strategies \citep{Ge2018Experimental}. 
In 1993, the first experiment with fully ACC-equipped vehicles was tested for string stability under the constant gap policy \citep{Ioannou1993Intelligent}.
More recently, a study on a circular track with 22 vehicles  demonstrates that traffic oscillations can be dampened by adjusting the speed of a single controlled vehicle \citep{Stern2018}.
In contrast, an empirical experiment with a platoon of seven (different) commercially available vehicles classified as SAE Level 2 showed that the platoon of cars in ACC mode amplifies perturbations \citep{Knoop2019Platoon}.

Overall, the current literature does not allow one to compare the effectiveness of multiple (C)ACC strategies because each study relies on different assumptions on the formation of stop-and-go waves. Further, most of the studies are focused on determining whether a (C)ACC strategy is string stable or unstable. This binary classification does not allow quantitative comparisons among strategies. 
This paper attempts to fill this gap by defining two quantitative measures to compare the oscillation growth in Section \ref{sec:measures}. Moreover, we compare the effectiveness of multiple (constant time gap strategies) for different types of vehicles, depending on their automation and connectivity capabilities, defined in Section \ref{sec:vehicles}. This systematic framework and comparison allow quantifying the effect of different (C)ACC strategies on the stop-and-go waves at low MPRs and suggesting which technologies are most effective in mitigating the oscillations.

\section{Methodology}

\subsection{Stop-and-go formation with noise}\label{secAlcoba}

\begin{figure}
\centering
     \includegraphics[width=0.6\textwidth]{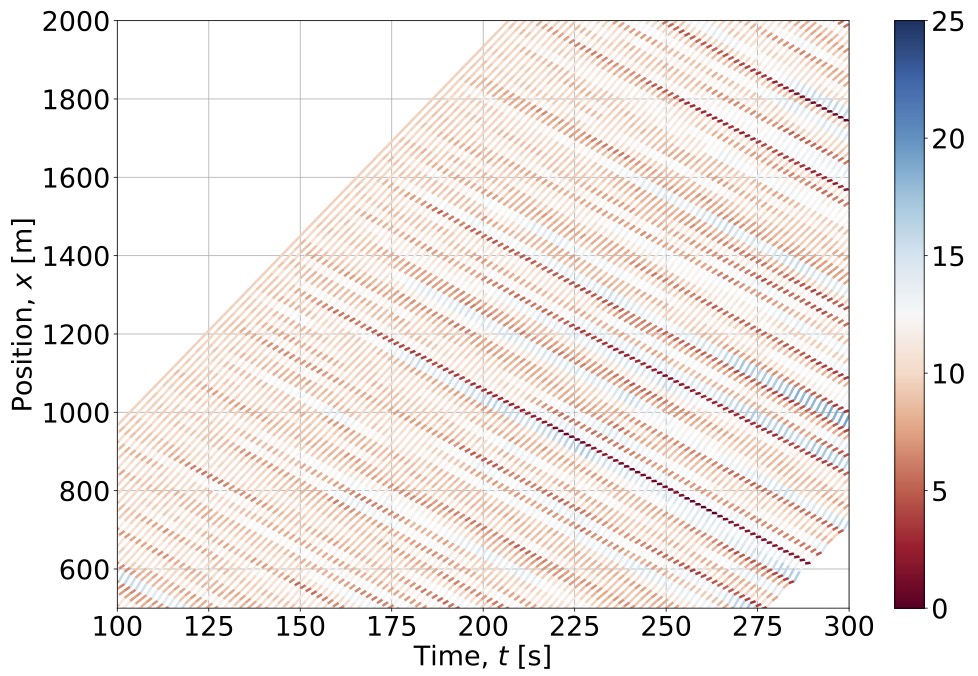}
    \caption{ Trajectories of vehicles with an initial spacing of 22m, where he color indicates the speed [m/s]. }\label{fig:stopandgo}
\end{figure}

In this study, we assume the stop-and-go formation originates from the (human) perception error on the spacing to the leader, as proposed in \citep{Alcoba2019LWR}. The NSCF model \citep{Newell2002} is extended by introducing a stochastic noise term in the speed-spacing relationship as $\hat V(s)= V(s) + \hat \sigma(v)  \dot W$, i.e.,
\begin{align}
    \hat V(s) = \min \{ u_0, \frac{s  - s_j}{\tau}  \} + \hat \sigma(v) \dot W,\label{eq/vAlcoba}
\end{align}
where $n$ indicates the vehicle number, $u_0$ is the free-flow speed, $\hat \sigma(v)$ is a speed-dependent model parameter, $ \dot W$ is the random term (white noise), which follows a normal distribution with zero mean and unit variance, and the other parameters were defined in Section \ref{sec:intro}. 
The calibration of $\hat \sigma(v)$ \citep{Alcoba2019LWR} was based on empirical data \citep{Jiang2014}. 
We assume a constant $\hat \sigma$ in this study for simplicity. 
Further,  we assume that the free-flow speed is never surpassed, and the vehicles do not travel backwards, i.e., the speed of vehicle $n$ is
\begin{align}
    v(t + \tau, n) = \min \{ u_0, \max \{ \hat V(s_d(t,n) ), 0 \}\}.\label{eq:HV}
\end{align}
An example of 100 vehicles' trajectories following a leader travelling at constant speed of 9.66 m/s is presented in Fig \ref{fig:stopandgo}. 

\subsection{Assumptions on different intelligent vehicle types}\label{sec:vehicles}

We consider different types of vehicles depending on their data acquisition capabilities. 
Based on the technology available on the vehicles, their speed will be determined by a chosen desired spacing $s_d(t,n)$ and  the $\hat \sigma$ parameter with Equation \ref{eq:HV}. 
The idea is to evaluate which control and data collection mechanisms are more effective in damping stop-and-go waves at low MPRs of intelligent vehicles.

We adopt the following terminology to classify the different types of vehicles: 
\begin{itemize}
    \item HVs are neither automated nor connected and measure the gap to the leader with an error. They adapt the speed to the measured spacing $s(t,n)$ to their leader.
    \item Vehicles that have some level of automation (i.e., can measure the spacing to their leader accurately) are referred to as automated vehicles (AVs). For these vehicles, we assume $\hat \sigma=0$.
    \item Vehicles with multi-anticipation detection technology are referred to as MAVs. These vehicles are assumed to detect the location of the first and the second leader without errors. Their desired spacing is the average between both leaders.
    \item Connected vehicles (CVs) are equipped with communication devices and can exchange information. We will consider both partial connectivity (PC), where PCVs exchange information on their measured spacing through V2V, and full connectivity (FC), where FCVs receive information from the infrastructure through V2I on the average density in the road.
    \item When a vehicle is automated and exchanges information, we refer to it as a connected and automated vehicle. In particular, we will refer to them as PCAV or FCAV depending on the communication technology
\end{itemize} 
The names and abbreviations of the seven types of vehicles considered are presented in Table \ref{tab1}. 

From (\ref{eq/vAlcoba}) and (\ref{eq:HV}), for vehicles with measurement errors, i.e., HVs, PCV, and FCV, the speed follows
\begin{align*}
v(t + \tau, n) = \min \{ u_0, \max \{ \min \{ u_0, \frac{s_d(t,n)  - s_j}{\tau}  \} + \hat \sigma \dot W , 0 \}\}.
\end{align*}
In contrast, the speed of intelligent vehicles without measurement error (i.e., $\hat \sigma =0$), the speed is determined as
\begin{align*}
  v(t + \tau, n) = \min \{ u_0,  \frac{s_d(t,n) - s_j}{\tau}  \},
\end{align*}
where $s_d(t,n)$ is determined depending on the technological equipent of the vehicle.
For example, PCAVs exchange information about the spacing between connected vehicles and cooperatively adjust their speed to match the average spacing following, i.e.,
\begin{align}
 s_d(t,n) = \frac{1}{n_c}\sum_{\forall c}{s(t,c)},\label{eq:PCAV}
\end{align}
where  $n_c$ denotes the total number of PCAVs present and $s(t,c)$ is the spacing measured by each PCAV vehicle. It should be noted that this particular strategy, although intuitive, has not been previously proposed in the literature. Alternatively, PCAVs could opt for a more cautious approach and select the speed corresponding to the smallest measured spacing $s(t,c)$. 
In the cases where these partially connected vehicles are not automated, i.e., PCVs, the desired spacing would be the same, but speed from (\ref{eq:PCAV}) will be modified to introduce the measurement error (noise) and non-negativity constraints, similar to (\ref{eq:HV}).

\begin{table}
\caption{\small Types of (Intelligent) Vehicles Considered}
\begin{center}
\begin{tabular}{|l|c|c|}
\hline 
\textbf{Vehicle type (Abbreviation)}& \textbf{\textit{$s_d(t,n)$}} & \textbf{\textit{$\hat \sigma$}} \\
\hline
Human-driven (HV) & $s(t,n) $ & $0.25$ \\
Automated  (AV) & $s(t,n)$  & $0$ \\
Multi-anticipation automated  (MAV) & $\frac{s(t,n) + s(t,n-1)}{2} $ & $0$ \\
Partially connected (PCV) & $\dfrac{\sum_{\forall c} s(t,c)}{n_c} $  & $0.25$\\
Partially connected and automated  (PCAV) & $\dfrac{\sum_{\forall c} s(t,c)}{n_c} $  & $0$ \\
Fully connected (FCV) & $\bar s  $  & $0.25$ \\
Fully connected and automated  (FCAV) & $ \bar s $  & $0$ \\
\hline
\end{tabular}
\label{tab1}
\end{center}
\end{table}

As aforementioned, we assume that under full connectivity, the average density, $k(t)$, of the surrounding traffic is shared through V2I to the F(A)CVs. Thus, the desired spacing is $s_d(t,n) = \bar s = \frac{1}{k(t)}$. 
For MAVs, the desired spacing is the average spacing to their first and second leader, as shown in Table \ref{tab1}.

It is worth noting that there is a consensus in the traffic flow community that AVs, and CAVs will have a different time gap than HVs, modifying the road capacity. While we acknowledge this fact, this paper focuses on the oscillation impact, and thus we assume the same time gap for all vehicles.

\subsection{Metrics to study the oscillation propagation}\label{sec:measures}

To accurately compare the impact of different intelligent vehicles presented in Section \ref{sec:vehicles} on the stop-and-go traffic, it is important to define relevant traffic flow metrics to compare the oscillations' growth. We propose two metrics:

\begin{itemize}
\item The first metric describes the oscillation propagation from vehicle to vehicle. We consider the standard deviation of the speed for each vehicle in the platoon over a certain time period. This standard deviation was measured in empirical studies \citep{Jiang2014} for HVs platoons and was shown to increase with the length of the platoon proportional to $\sqrt n$ for the stochastic LWR \citep{Alcoba2019LWR}. To obtain this metric, we will rely on simulations of the lead-vehicle problem with a constant speed of the leader and a platoon of $N$ vehicles.
\item The second metric describes the oscillation propagation in time under the assumption of an infinite platoon. This can be measured by modeling a circular road where $L$ is the length, and $N$ is the number of vehicles in the ring road. Then, the standard deviation of the speed at each time over all vehicles $N$ can be obtained.
\end{itemize}

The oscillation growth will depend on the position of the intelligent vehicles between the other HVs. Therefore,  to quantitatively compare the evolution of the standard deviation of speed for a given MPR, we propose relying on Monte Carlo Simulations (MCS). In each simulation run, the location of the intelligent vehicle is randomly assigned within the platoon. Then, the expected oscillation growth will be obtained from the mean of all MCS.

\section{Numerical Results}

%\subsection{Set up}
%\subsection{Simulation results}

In this section, we analyze the impact of different vehicles on stop-and-go propagation based on the metrics presented in Section \ref{sec:measures}.  
We assume the traffic flow parameters to be the same for all vehicles: free flow speed, $u_0 = 25$ m/s, jam spacing, $s_j=7.5$ m, and time gap, $\tau=$1.5 s.%, and  $\hat   \sigma=0.25$ m/s.

\subsection{Oscillations propagation vehicle to vehicle}\label{sec:open}

In this section, we present the results of the first metric to evaluate the oscillation propagation from vehicle to vehicle. 
We  consider a linear open road with a platoon of $N$ vehicles, where the leader has a deterministic and constant speed. 

\begin{figure}
\centering
    \begin{subfigure}{0.5\textwidth}
	   \includegraphics[width=\textwidth]{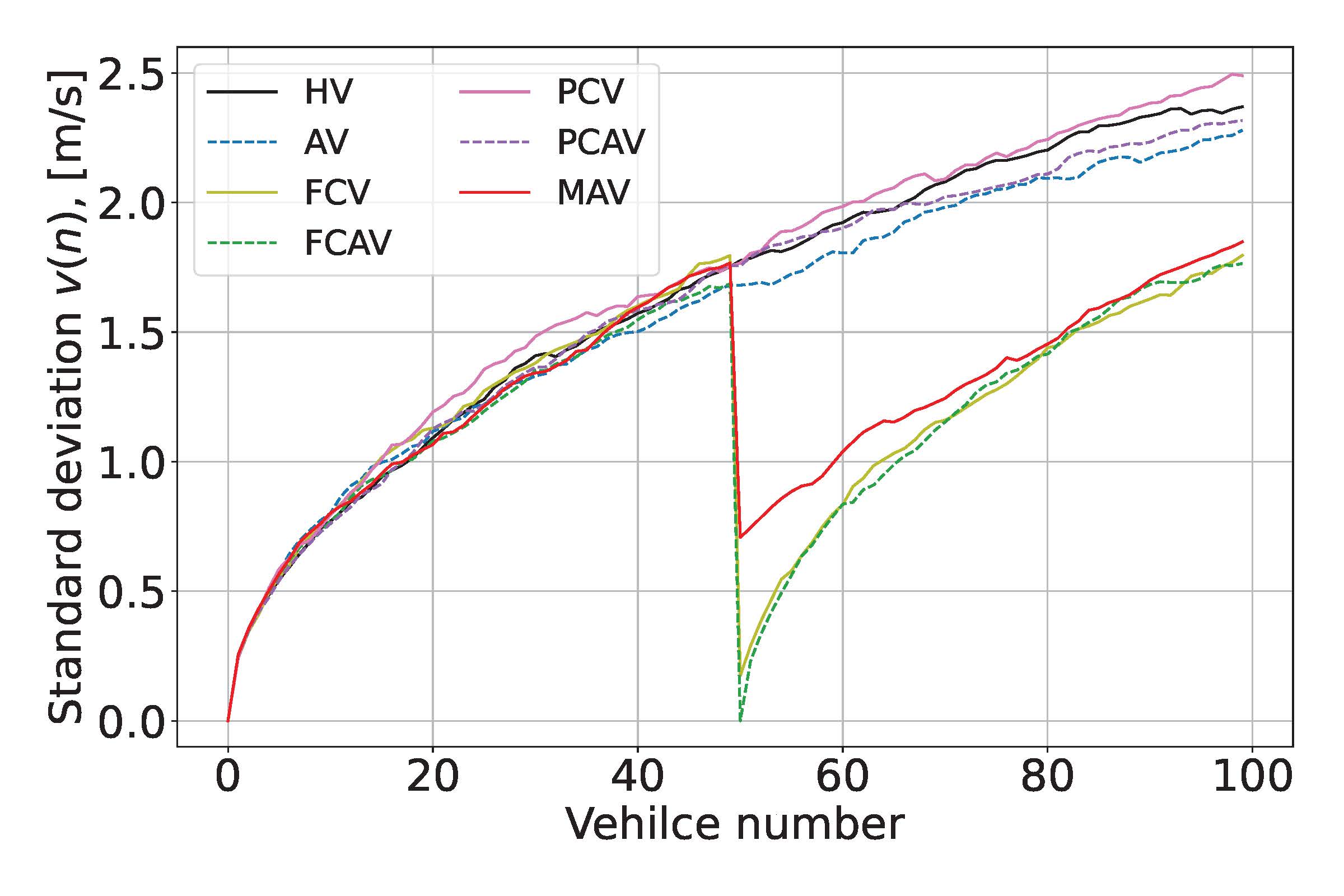}
	   \caption{\small } \label{fig:std-N1}
    \end{subfigure}
    \\
     \begin{subfigure}{0.5\textwidth}
            \includegraphics[width=\textwidth]{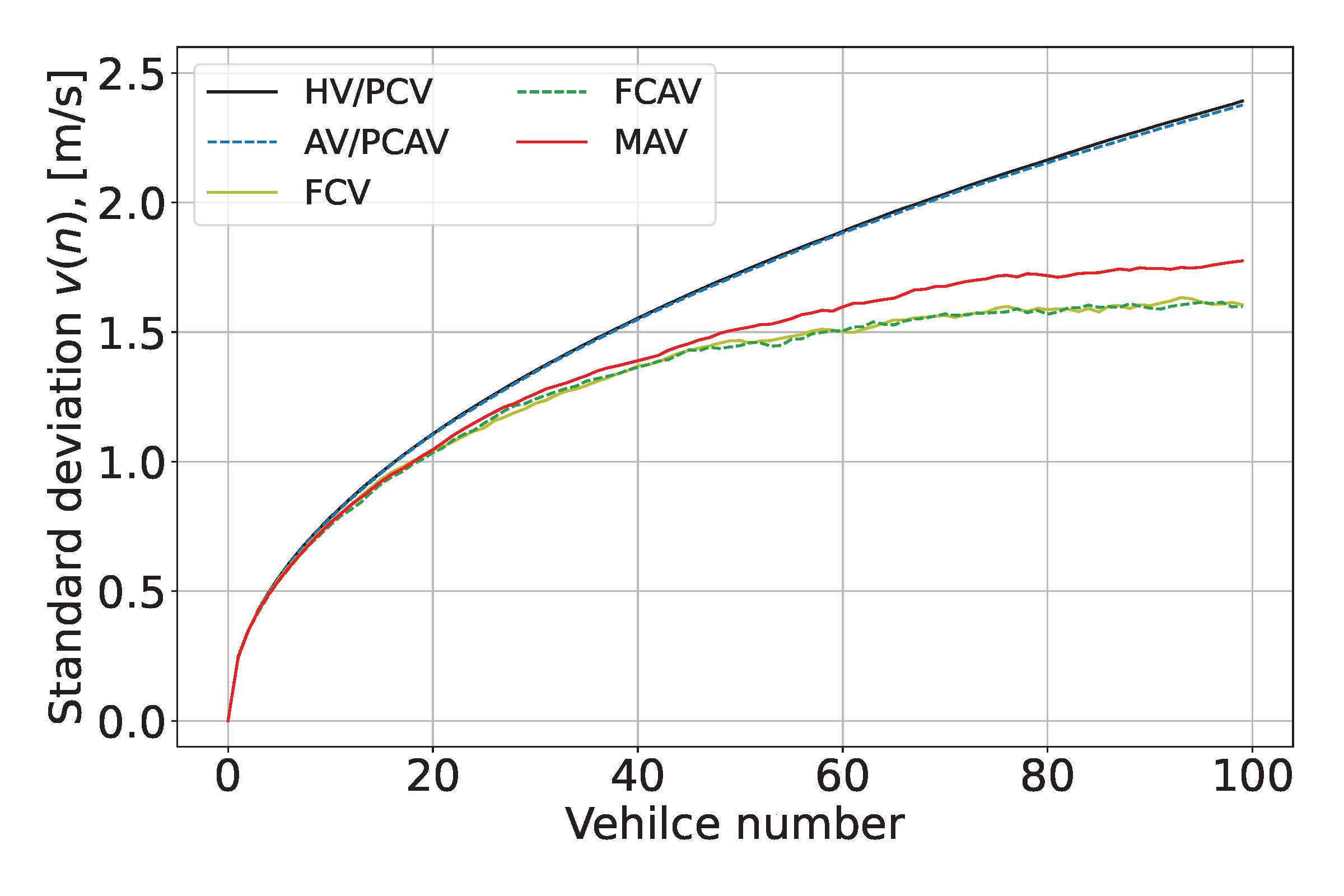}
            \caption{\small} \label{fig:std-n}
    \end{subfigure}
    \caption{ Speed variation of each vehicle in a platoon with one intelligent vehicle. (a) In position \#50, (b) average over 500 MCS with a random location of the intelligent vehicle.}
\end{figure}

First, let us consider a platoon of $N=100$ where a single vehicle in the middle of the platoon (number 50) is equipped with (C)ACC technology. In this scenario, partial connectivity does not bring any additional benefit because the vehicle is ``alone'' and cannot communicate with others. Thus, the PCV behaves as an HV, and the PCAV behaves as an AV.
Fig. \ref{fig:std-N1} shows how a vehicle equipped with full connectivity and automation (FCAV) completely eliminates the oscillation growth and acts as a barrier between platoons. Similarly, the FCV is capable of almost fully eliminating the oscillation. We can conclude that FC(A)Vs are effectively acting as a new platoon leader. The MAV also significantly reduces the oscillation but does not eliminate it completely.
In contrast, automation alone (AV and PCAV) presents minor benefits compared to a platoon with only HVs.

To better quantify the impacts of each type of technology, we study 500 MCS for different (random) locations of the intelligent vehicle (Fig. \ref{fig:std-n}). The results confirm the general observations from Fig. \ref{fig:std-N1}. The expected oscillations for vehicle \# 100 in the platoon are reduced by almost 20\% when a single MAV is introduced and by more than 25\% when a vehicle with full connectivity is present.

\begin{figure}
\centering
    \begin{subfigure}{0.5\textwidth}
	   \includegraphics[width=\textwidth]{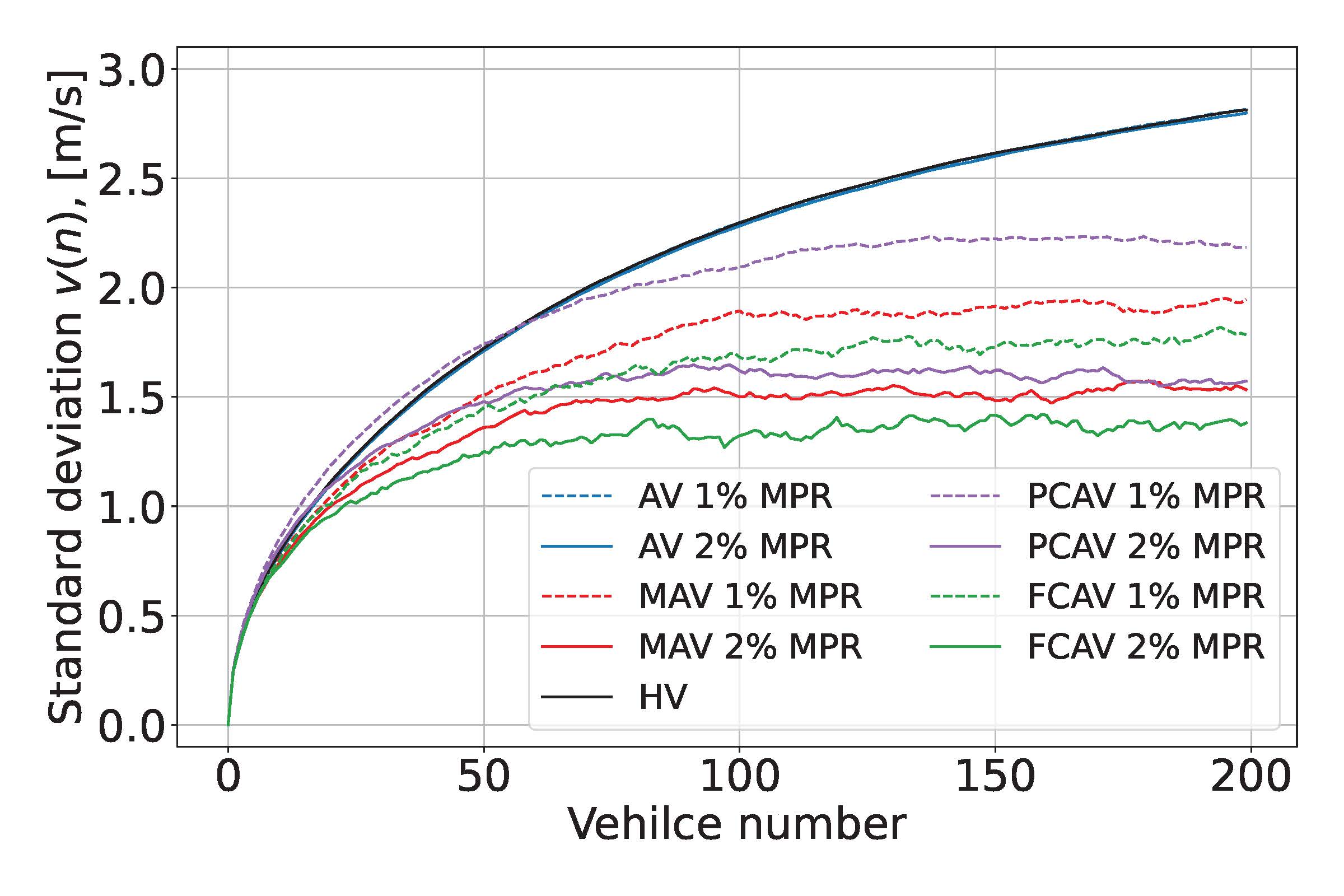}
	   %\caption{\small } \label{fig:std-n-MPR}
    \end{subfigure}
    \caption{  Variation of speed of each vehicle in the platoon under different MPR. Average over 250 MCS.}\label{fig:std-n-MPR}
\end{figure}

In the following, we consider a larger platoon ($N=200$) and analyze the impact of an increased MPR for AVs, PCAVs, FCAVs, and MAVs (Fig. \ref{fig:std-n-MPR}). In a platoon of 200 vehicles, partial connectivity may have an impact even for MPR=1\%. 
Again,  the impact of introducing AVs is negligible. 

At 1\% MPR the speed standard deviation of the last vehicle shows reductions of 15\% for PCAVs, 25\% for MAVs, and more than 30\% for FCAVs. 
When the MPR increases to 2\%, the differences between PCAVs and MAVs become less significant, with both reducing the standard deviation of the last vehicle by approximately 40\%. Remarkably, FCAVs achieve a significant reduction, lowering the speed standard deviation to 45\%.  It is worth noting that for MPR 2\% the standard deviation only increases from vehicle to vehicle until approximately vehicle \#80 in the platoon. 
All subsequent vehicles experience similar standard deviations, indicating that even with very low MPR, the increase in oscillations can already be effectively mitigated with MAVs, and vehicles with some level of connectivity.

\subsection{Ring road: Propagation of oscillations over time}\label{sec:ring}

When considering the oscillation propagation over time in a ring road, we are effectively analyzing an infinite platoon of vehicles.

\begin{figure}
\centering
     \includegraphics[width=0.8\textwidth]{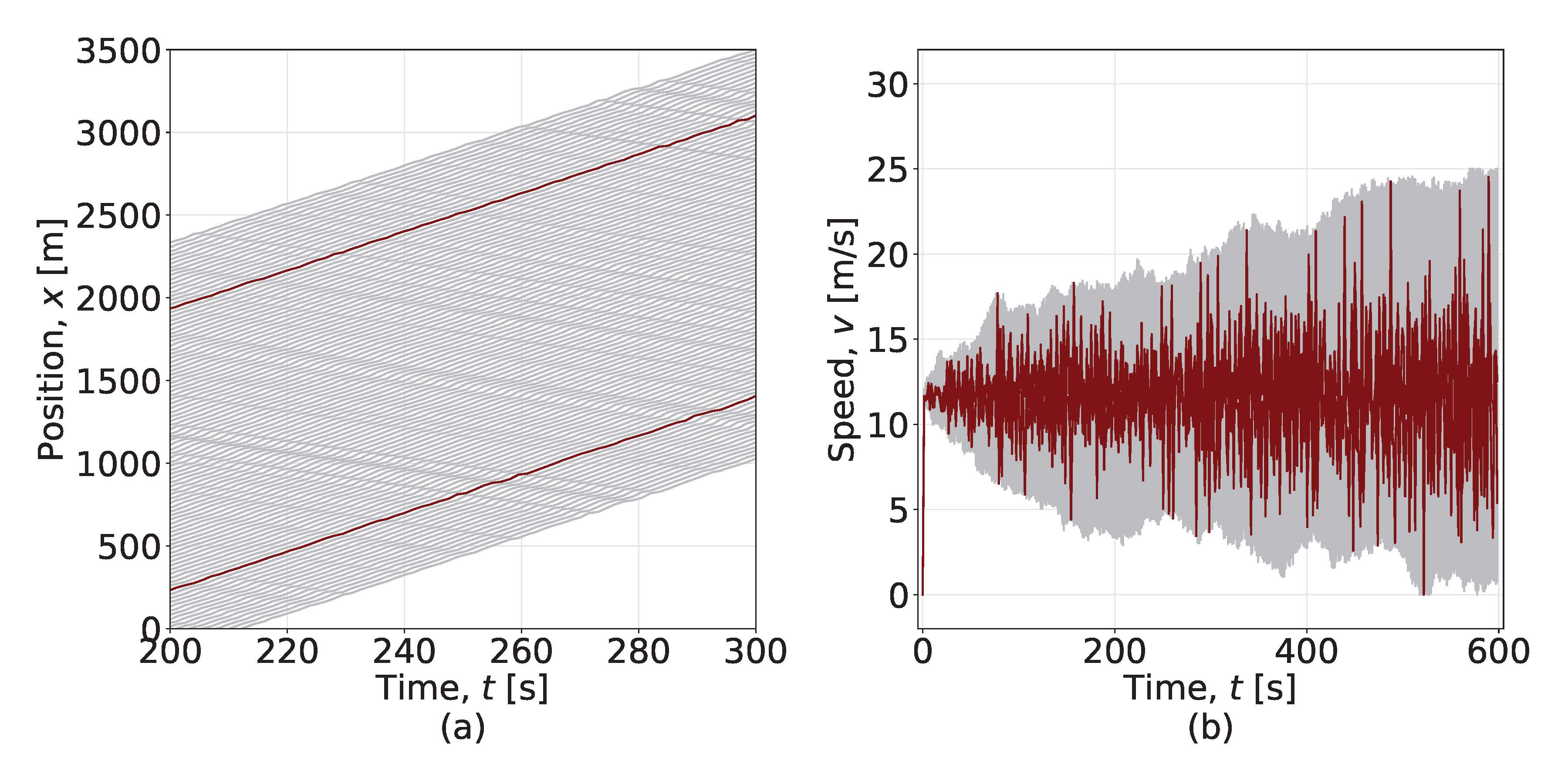}
    \caption{ Single simulation run on a ring road of $L=2.5$ km with MPR=2\% of Automated Vehicles (AVs). (a) Trajectories (b) The speed evolution  in m/s. }\label{fig:simAV}
\end{figure}

\begin{figure}
\centering
     \includegraphics[width=0.8\textwidth]{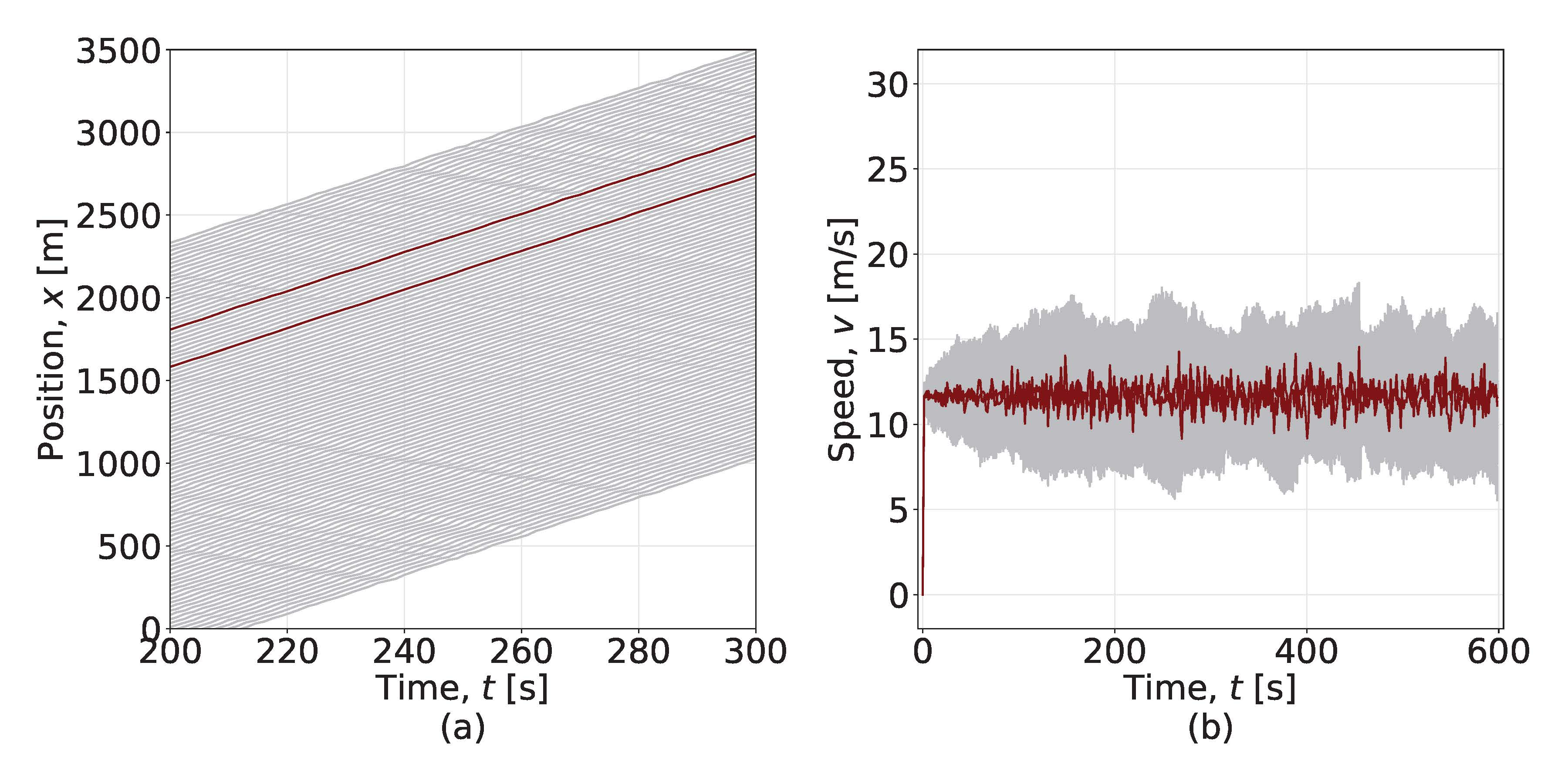}
    \caption{ Single simulation run on a ring road of $L=2.5$ km with MPR=2\% of Multi-anticipation automated vehicles (MAVs). (a) Trajectories (b) The speed evolution  in m/s.  }\label{fig:simMAV}
\end{figure}

\begin{figure}
\centering
     \includegraphics[width=0.8\textwidth]{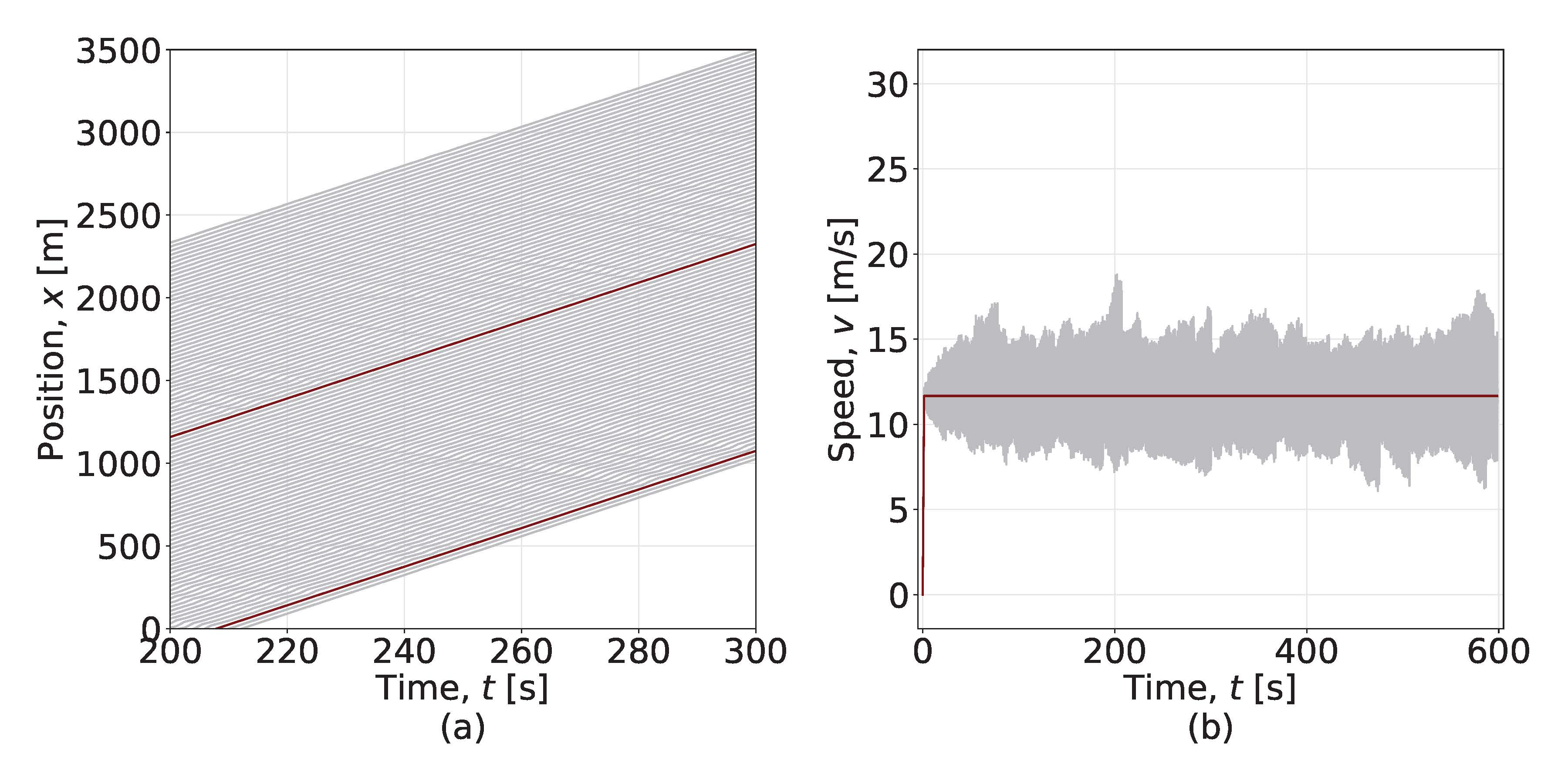}
    \caption{ Single simulation run on a ring road of $L=2.5$ km with MPR=2\% of Fully connected and automated vehicles (FCAVs). (a) Trajectories (b) The speed evolution  in m/s. }\label{fig:simFCAV}
\end{figure}

First, we present some qualitative results with a platoon of 100 vehicles on a ring road of $L=2.5$ km.
Due to space limitations, we only present three cases to qualitatively compare the impacts of low (2\%) MPR for AVs, MAVs and FCAVs in Fig. \ref{fig:simAV}, \ref{fig:simMAV}, and \ref{fig:simFCAV}, respectively. The gray lines are for data obtained from HV and red lines correspond to intelligent vehicles. From Fig. \ref{fig:simAV}a, the AVs cannot avoid the stop-and-go formation, and multiple shock waves are observed in the trajectories. After less than 10 min of simulation, the vehicles' speeds oscillate between 0 and $u_0$ m/s (Fig. \ref{fig:simAV}b). In contrast, when the vehicles are equipped with multi-vehicle anticipation technology, the oscillations are significantly reduced (Fig. \ref{fig:simMAV}).
Similar results are obtained from the scenario with V2I, where the FCAVs' (time-independent) speed is determined with $s_d(t,n) = \frac{L}{N}=25$ m.
Since these scenarios have different locations of intelligent vehicles on the ring road, one cannot quantitatively compare the results.

%\textcolor{red}{}

\begin{figure}
\centering
    \begin{subfigure}{0.37\textwidth}
	   \includegraphics[width=\textwidth]{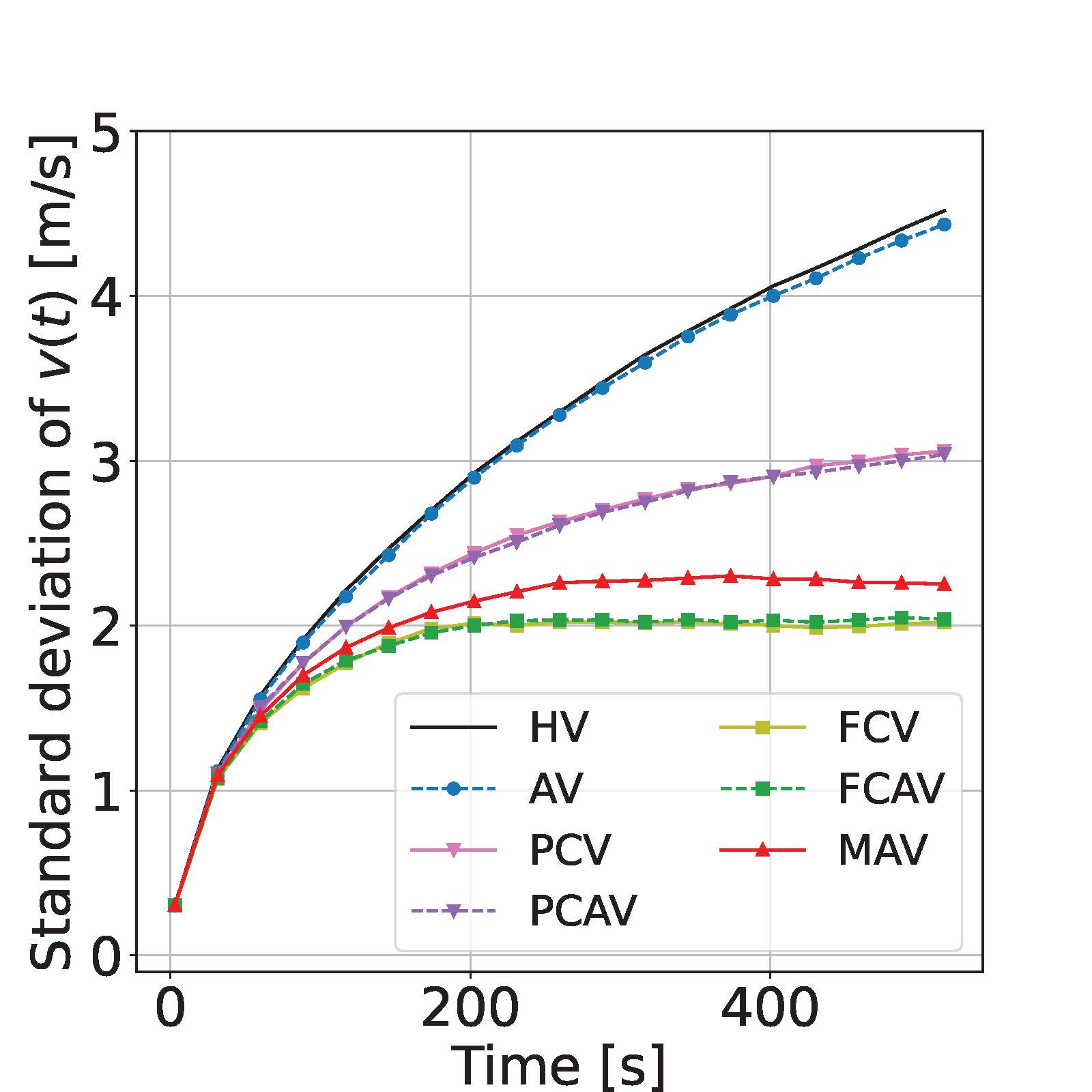}
	   \caption{  MPR of 1\% } \label{fig:1mpr}
    \end{subfigure}
     \begin{subfigure}{0.37\textwidth}
            \includegraphics[width=\textwidth]{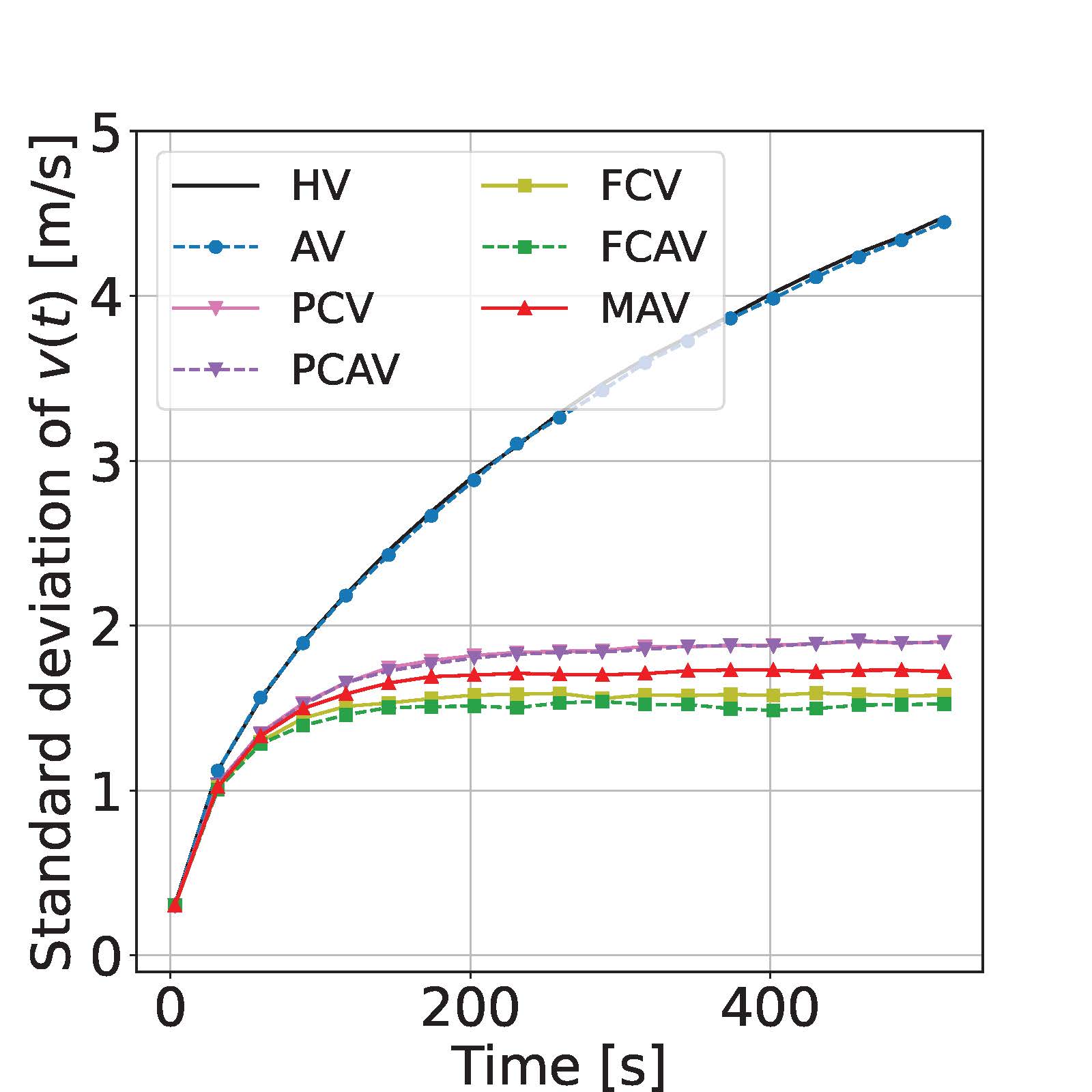}
            \caption{ MPR of 2\%}
    \end{subfigure}
    \begin{subfigure}{0.37\textwidth}
	   \includegraphics[width=\textwidth]{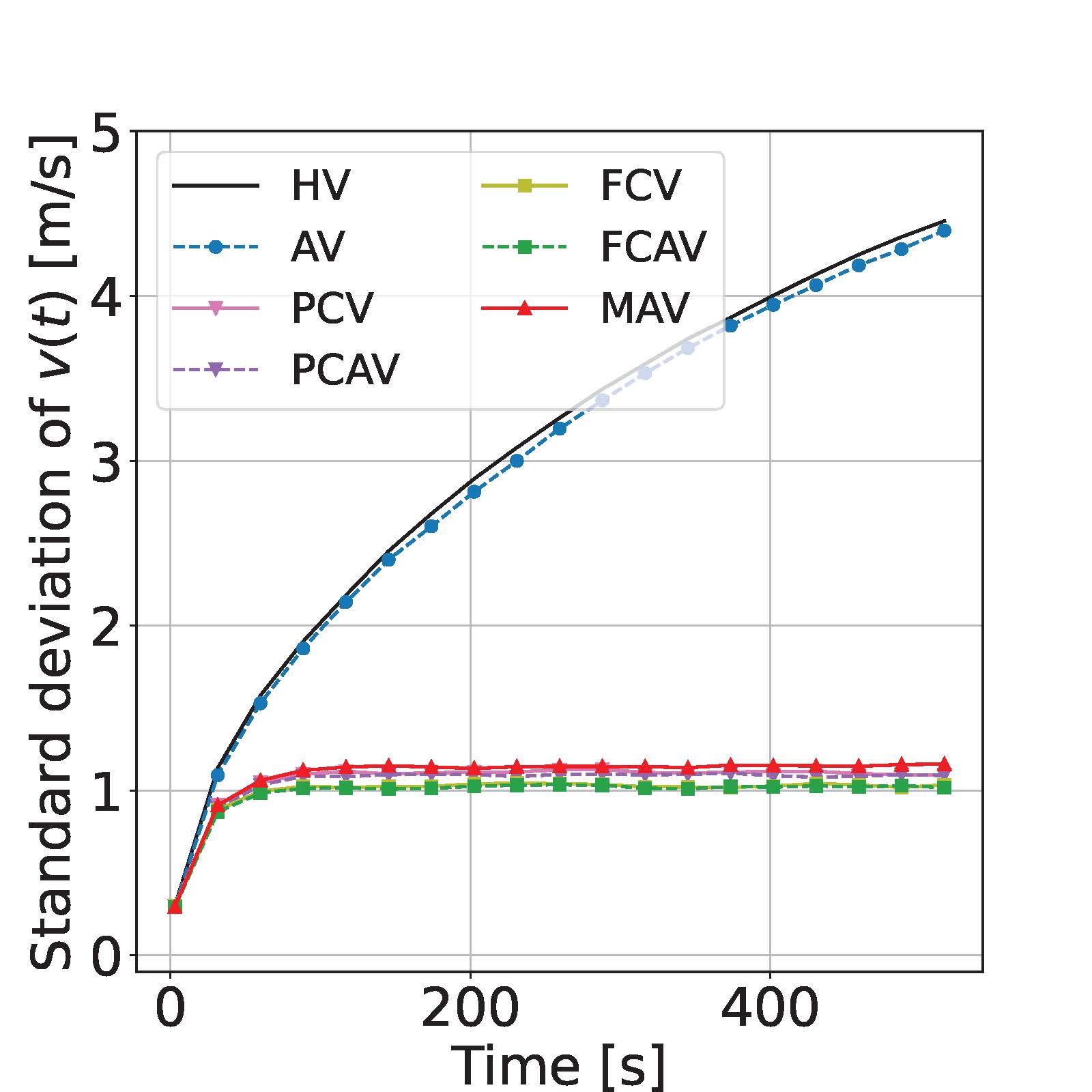}
	   \caption{  MPR of 5\% } 
    \end{subfigure}
     \begin{subfigure}{0.37\textwidth}
            \includegraphics[width=\textwidth]{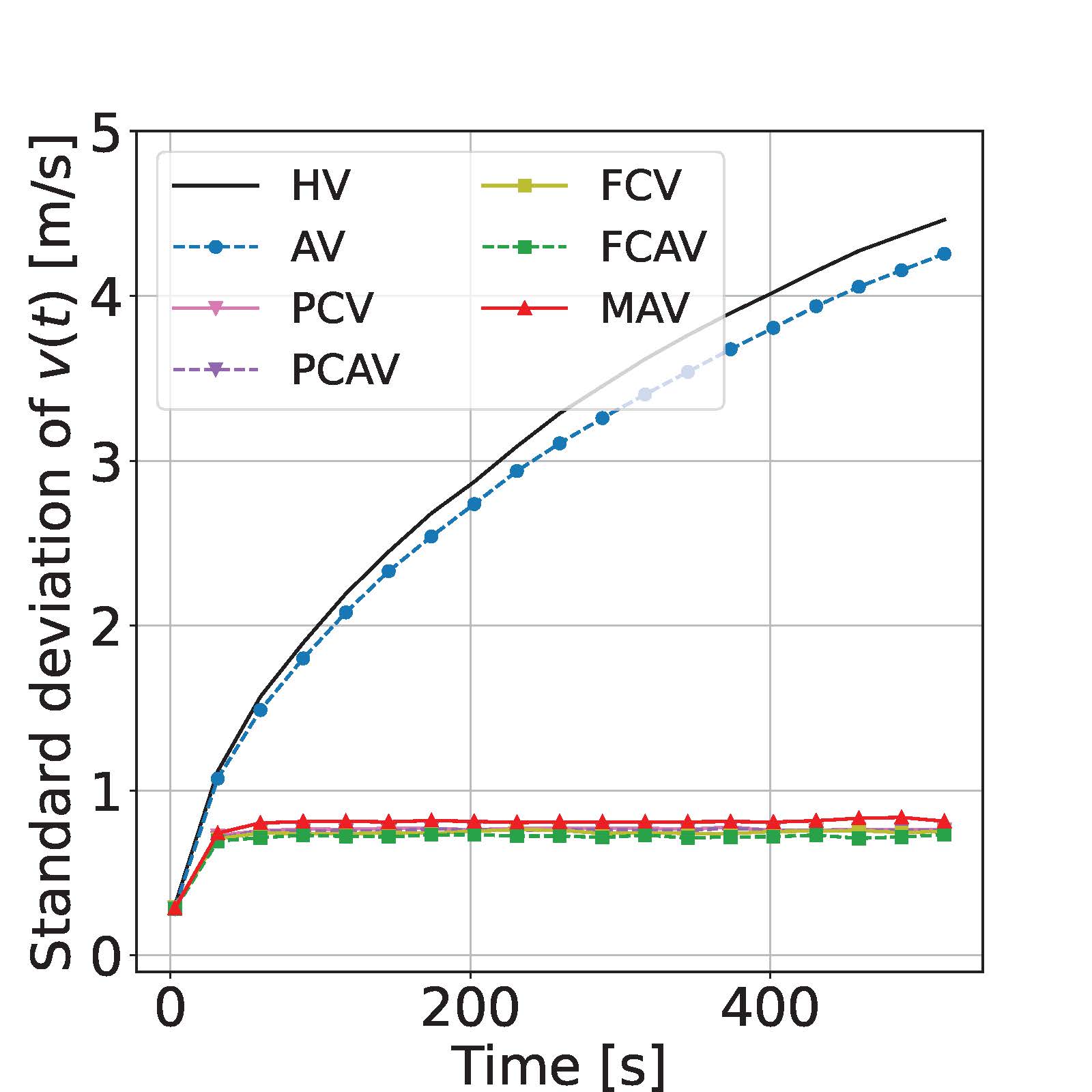}
            \caption{ MPR of 10\%}
    \end{subfigure}
    \caption{  Time evolution of the expected speed standard deviation across vehicles on ring road of $L= 6$ km (100 MCS), for different vehicle types and varying MPRs.}\label{fig:std-t}
\end{figure}

In the following, we discuss the results obtained from the average standard deviation for 100 MCS and all vehicle types. In order to evaluate the impact of partial connectivity at 1\% MPR, we model 200 vehicles in a ring road of 6 km.
The overall observations from the numerical results in Fig. \ref{fig:std-t} are consistent with the results in Section \ref{sec:open}. While the presence of AVs does not dampen the oscillations, full connectivity is the best way to mitigate stop-and-go at very low MPRs. 

For FC(A)Vs and MAVs, even a 1\% MPR is able to stabilize the oscillations around 2 m/s and 2.2 m/s, respectively. The fact that MAVs are effective at such MPRs is promising because vehicles do not rely on communication technology to reduce traffic oscillations.
Fig. \ref{fig:std-t}c,d show that MPRs of 5\% (and larger) most intelligent vehicles are able to limit the growth of the oscillations, keeping them at $\sim$ 1 m/s for 5\% MPR and slightly below for 10\% MPR. Our extensive numerical results show that this conclusions (for MPRs $\geq$ 5\%) are only slightly dependent on the ring road length as long as the density is kept constant. We observe slightly lower standard deviations for shorter ring roads.

\section{Discussion and conclusions}

The mitigation of stop-and-go waves through (cooperative) adaptive cruise control (C)ACC has gained interest in the literature. These studies aim to propose sting stable (C)ACC. In general, each study relies on a certain assumption regarding the formation and propagation of oscillations. Thus, the current body of literature hinders the comparison of efficacy among different string-stable (C)ACC strategies. 
This study defines two quantitative measures to compare oscillation growth under various conditions. Additionally, we compare multiple (constant time gap) (C)ACC strategies' effectiveness based on the automation and connectivity capabilities of the vehicles. 

As expected, the highest benefits are obtained when full connectivity is available, and the vehicles receive information about the average density (or equivalently a desired average speed). 
The numerical results show that AVs cannot eliminate stop-and-go at MPR$\leq$10\%. These results contradict previous empirical experiments \citep{Stern2018}, which suggest that a single vehicle can eliminate stop-and-go on a ring road with 22 vehicles (i.e., MPR of 4.5\%). However, their AV setup did not consider a constant time gap strategy, and their AV may act more similarly to other vehicle strategies considered in this paper.
Interestingly, new technological advances like the multi-vehicle anticipation measurement system of automated vehicles may have comparable results to full connectivity even at MPRs 1\%. In this study, we assume that the MAVs are equipped with technology to sensor the location of the second leader. However, some studies suggest that this technology could measure the distance up to the third leader (albeit with increasing error measurements) \citep{Dona2022Multianticipation}. It would be interesting to evaluate whether adding information on the third leader would further improve the oscillation reduction.

In the future, we are interested in performing a sensitivity analysis on how the oscillations grow depending on $\hat \sigma$, which has been assumed constant in this work, and in calibrating and validating the underlying stochastic car-following model considered in this paper. 
Furthermore, the results and discussions presented in this paper are based on the average of MCS and revolve around the expected impact of different intelligent vehicles given a low MPR. However, it is possible that the location of the vehicle(s) within the platoon significantly impacts the results. Thus, a remaining practical research question is whether the oscillation growth can be more effectively dampened through a sequence of consecutive intelligent vehicles or by evenly distributing them throughout the platoon. While it seems intuitive that distributing MAVs will lead to better results, the same reasoning may not be applicable when considering other analyzed intelligent vehicles. Thus, future research should consider the impacts of the location of these vehicles within the platoon.

\section*{Copyright}

Copyright © 2023 IEEE. All rights reserved.
To be presented at ITSC in Bilabo: https://2023.ieee-itsc.org/detailed-program/
The link will be updated.

\small

\pdfbookmark[1]{References}{references}
%\bibliographystyle{elsart-harv}
%\bibliographystyle{apalike}
%\bibliography{references}
%\bibliographystyle{elsart-harv}
\bibliographystyle{apalike}

\bibliography{references_all}

\end {document}